\documentclass[11pt]{article}
\usepackage{amssymb}

\begin{document}
\def\bee{\begin{equation}}
\def\eeq{\end{equation}}
\def\kar{K\'a\-roly\-h\'azy} 
\def\schr{{Schr\"o\-din\-ger}}
\def\uc{uncertainty\  }
\def\ucs{uncertainties\  }
\def\st{space-time\  }
\def\gcm{\mbox{~{\rm g/cm}$^3$}} 
\def\gmb{\mbox{$(g_{\mu\nu})_\beta$}}
\def\dlt{\mbox{$\Delta_L T$ }}
\def\dlts{\mbox{$(\Delta_L T)_{\rm syn}$ }}
\def\sec{\mbox{\rm sec}}
\def\vr{\mbox{$|V\rangle$}}
\def\txt{\mbox{$\widehat t(\mathbf x,t)$}}
\def\txvt{\mbox{$\widehat t(\mathbf x',t)$}}
\def\txtv{\mbox{$\widehat t(\mathbf x,t')$}}
\def\taxt{\mbox{$\widehat \tau(\mathbf x,t)$}}
\def\taxvt{\mbox{$\widehat \tau(\mathbf x',t)$}}
\def\taxtv{\mbox{$\widehat \tau(\mathbf x,t')$}}

\title{\large \bf A TENTATIVE EXPRESSION OF THE 
 K\'AROLYH\'AZY UNCERTAINTY OF THE SPACE-TIME STRUCTURE THROUGH
VACUUM SPREADS IN QUANTUM GRAVITY}
 
\author{Andor Frenkel\\KFKI -- Research Institute for\\
Particle and Nuclear Physics, Budapest, Hungary
}
\date{}
\maketitle
\begin{abstract} 
\noindent
In the existing expositions of the K\'arolyh\'azy model, quantum
mechanical uncertainties are mimicked by classical spreads. It
is shown how to express those uncertainties through entities of
the future unified theory of general relativity and quantum theory.
\end{abstract}

\section{Introduction}
\label{s:1}
In his pioneering work \cite{1, 2} on the stochastic modification of
the Schr\"odinger time evolution, known as the K model (for a
recent review, see \cite{3}), \kar\  relates the loss of coherence and
the breakdown of the superposition principle to the \uc of the
Einsteinian space-time structure, caused by the quantum mechanical
\ucs of the position and of the momentum of material objects.
The K model leads to sound results concerning the occurrence (or
the absence) of the breakdown of the superposition principle for
various physical systems. It should be noted that in obtaining
those results there is no room for maneuvers, because there are
no free parameters in the K model. However, the theoretical
reliability of the results is weakened by the fact that in the
calculations the quantum mechanical \uc of the space-time
structure has been replaced by a classical spread. In \cite{1, 2}
(see also \cite{4, 5}) the single space-time $S$, with a quantum
mechanical \uc in its structure, has been mimicked by a
stochastic set $\{S_\beta\}$ of classical space-times $S_\beta$,
with appropriately chosen metric tensors $(g_{\mu\nu}(\mathbf x,
t))_\beta$. Instead of propagating on the single but ``hazy''
space-time $S$, the quantum mechanical wave function had to
propagate on the concise, but different from each other
space-times $S_\beta$, and the spread of the relative phases of
the wave function over the set $\{S_\beta\}$ was supposed to be
equal to the amount of the quantum mechanical \uc of the
relative phase, induced by the uncertain structure of the single
space-time $S$. In \cite{6, 7} an other stochastic space-time model,
leading to the same result as the original model, has been used.
In this second model $g_{\mu\nu}$ remains concisely Minkowskian, but
the moments of time are randomized via the introduction of an
appropriate random set $\{ t_\beta(\mathbf x, t)\}$ of moments of
time $t_\beta(\mathbf x, t)$.

\kar\  was well aware of the shortcomings of such \st models. In \cite{4}
he wrote: ``To avoid possible misunderstanding, we would like to
stress that \dots\ no physical significance should be attached
to the individual members of the family $\{ ( g_{\mu\nu})_\beta\}$. The
only role of the family is to provide us with a mathematical
model  of a single physical \st with smeared metric\dots''

Until recently, I thought that because of the lack of a unified
theory of general relativity and quantum theory, the use of a
classically stochastic \st model is inevitable. It turns out
that this is not so. As shown in the present paper, with the
help of two entities of the future unified theory 
of general relativity and quantum theory, 
one can obtain
the known results of the K model without relying on such a
space-time model. In addition, the calculations are simpler than
those carried out with the help of the \gmb's in \cite{2} (outlined
also in \cite{4, 5}). There, many expressions contain $\mathbf k$-space
Fourier sums or integrals. In the formulas for the physical
quantities these sums or integrals are convergent, but in some
intermediary expressions they diverge. In the present work the
calculations are carried out in $\mathbf x$-space, $\mathbf k$-space
integrals do not appear at all, and it is easy to see that the
claims that the K model needs a cutoff parameter to make
divergent $\mathbf k$-space integrals finite \cite{8} are not justified.

In Sections \ref{s:2} and \ref{s:3} two relations exhibiting the \uc of the
space-time structure are given and discussed. In Section~\ref{s:4} these
relations are expressed with the help of entities of the future
unified theory of general relativity and quantum theory. In
Sections~\ref{s:5} and \ref{s:6} a new derivation of the formulas for two basic
quantities of the K model --- the quantum mechanical spread
$\Delta_\Phi$ of the relative phases and the cell length $a_c$ ---
is presented.  The characterisation of quantum and of classical
behavior, and the transition between them are also discussed in
Section~\ref{s:6}. Section~\ref{s:7} is devoted to the presentation of the 
stochastically modified \schr\ evolution.
In Section~\ref{s:8} a few concluding
remarks are made. 

\section{Uncertainty in the Structure of the Einsteinian
Space-Time and in the Relative Phases of the Wave Functions}
\label{s:2}

\subsection{The Lower Bound of the Uncertainty of the
Length of Time Intervals}
\label{ss:2.1}

Investigating the \ucs of the Einsteinian \st structure induced
by the quantum mechanical \ucs in the position and in the
momentum of various quantum objects, \kar\  discovered that the
\uc of the length $T$ of a time interval has a lower bound \dlt.
The relation between $T$ and \dlt is simple:
\begin{equation}
\dlt \approx  T^{2/3}_P T^{1/3},
\label{eq:1}
\end{equation}
where
\begin{equation}
T_P = {\Lambda \over c} = \sqrt {G\hbar  \over c^5} = 5.3 \times
10^{-44} \sec
\label{eq:2}
\end{equation}
is the Planck time,
\begin{equation}
\Lambda = 1.6 \times 10^{-33}
\label{eq:3}
\end{equation}
being the Planck length and $G$ the constant of gravitation. 

The approximate equality sign $\approx $ in relation (\ref{eq:1}) and in
further equations takes into account that because of the absence
of a unified theory of general relativity and quantum theory,
numerical factors, unlike the $1/2$ in Heisenberg's relation
$\Delta x \cdot \Delta p \geqslant \hbar / 2$, could not be fixed. In
the context of the present paper such factors are unimportant.
Indeed, as we shall see, the value of the relevant parameter ---
of the cell length --- characterizing the coherence changes by
tens of orders of magnitude while going from quantum behavior to
classical behavior. Compared to this change, a shift by a factor
between $10^{-1}$ and $10$, or even between $10^{-2}$ and
$10^2$, is irrelevant. However, in a prospective experimental
search for the anomalous Brownian motion predicted by the K
model \cite{2, 4}, the said loose factors may cause problems.

Taking advantage of the regrettable looseness in the basic
relation (\ref{eq:1}), we shall often lump into the symbol $\approx $ known
but neglected factors, e.g.\ when rounding numerical values or
dropping $\pi$'s.

Two restrictions should be made concerning the applicability of relation~(\ref{eq:1}).

(i) In agreement with the fact that the K model is a model for
non-relativistic quantum mechanics, in relation (\ref{eq:1}) $T$ refers
to a time interval along a world line of a body slowly moving
(or standing) in a reference frame in which the 2.7$^\circ$~K
background radiation is isotropic.

(ii) When
\begin{equation}
T \lesssim T_P,
\label{eq:4}
\end{equation}
the very concept of space-time becomes questionable, and
relation (\ref{eq:1}) may lose its physical meaning. Therefore, this
relation should be applied only to time intervals for which
\begin{equation}
T \gg T_P. 
\label{eq:5}
\end{equation}

This restriction means that 
besides being a fundamental parameter of the K model, $T_P$ is also
a physical cutoff
parameter. It is the only inherent cutoff parameter in the 
model. It is needed in order to keep out from those very small
\st domains inside which the physical laws are not known, and
not in order to make divergent mathematical expressions finite.
Of course, similarly to regular non-relativistic quantum
mechanics, the predictions of the K model become unreliable (but
not divergent) when the realm of high energy particle physics
is reached, that is already for time intervals of the order of
$10^{-24}$~sec and for spatial distances of the order of
$10^{-13}$~cm, much larger than $T_P$ and $\Lambda$, respectively.

Notice that for $T = 1$~sec, \dlt is only of the order of
$10^{-29}$~sec. On the other hand, \dlt is an absolute,
inescapable lower bound. For a given $T$, it cannot be
diminished at the expense of some other quantity, like $\Delta
x$ at the expense of $\Delta p$ in Heisenberg's \uc relation. 

The extreme smallness of \dlt is due to the fact that when
deriving relation (\ref{eq:1}), only the basic laws of general relativity
and of quantum mechanics should be respected, all other
theoretical, as well as all practical limitations should be
ignored. As a result, the minimal time \uc of actual physical
processes of duration $T$ is much larger than the lower bound
\dlt for that value of $T$. \dlt could be reached only in
processes with bodies of irrealistically high density, although
this density is still much-much smaller than the Planck density
$\varrho_P = m_P / \Lambda^3 \approx  10^{94}\gcm$. Thus, the
existence of such irrealistic bodies would not contradict the
basic laws of general relativity and of quantum mechanics.

Only the existence and the order of magnitude of the lower bound
\dlt is exploited in the K model, the possibility of reaching it
is not necessary.

\subsection{The Uncertainty of the Space-Time Structure
and the Breakdown of the Superposition Principle}
\label{ss:2.2} 

Attention should be payed to the rather remarkable fact 
that all the physical quantities referring to a particular body
(like its mass, its velocity, etc.) dropped out from relation
(\ref{eq:1}), the relation between $T$ and \dlt involves only the
universal constant of nature $T_P = \sqrt{G\hbar / c^5}$.  Now,
a \st relation independent of any particular property of matter
can be, perhaps even must be attributed to \st itself, therefore
also to the empty space-time. Accordingly, \kar\ 
proposed to regard relation (\ref{eq:1}) as an expression of the
\uc of the structure of space-time. \dlt gives then the measure
of the limitation of the sharpness of the Einsteinian \st
structure, imposed by quantum mechanics. As shown in \cite{1, 2},
this tiny uncertainty of the \st structure induces \ucs in the
relative phases of the wave function of any isolated system, and
thereby limits, in return, the sharpness of the phase relations.
The amount of the \uc of the relative phases turns out to be
negligible in the case of microsystems, but it is large enough
to destroy the coherence of the wave function of a macrosystem,
in agreement with the observed breakdown of the superposition
principle. 

\section{The Structural Uncertainty of Synchronization}
\label{s:3}

The structural \uc \dlt of the length of the time intervals
along the $|\mathbf v| \ll c$ worldlines produces a structural
\uc of the same order of magnitude in the synchronization of the
times between two such worldlines \cite{6}. In order to see this, let
us consider, first in Minkowskian space-time, two $|\mathbf v| =
0$ worldlines $W_1$ and $W_2$ at a distance $r$ from each other.
Let the times along these worldlines be synchronized. A light
signal emitted on $W_1$ arrives back to $W_1$ from $W_2$ after a
time $2T$, where
\begin{equation}
T = {r \over c}.
\label{eq:6}
\end{equation}

In the \st of \kar,  the time interval of length $2T$ along $W_1$
has the structural \uc \dlt given by relation (\ref{eq:1}). (The
\ucs of $2T$ and of $T$ are of the same order of magnitude.)
Consequently, the moment of arrival of a signal to $W_2$ suffers
from the same \uc relative to the time along $W_1$. This means
that the \uc of the synchronization of the times along two
$|\mathbf v| = 0$ worldlines at a distance $r = cT$ from each
other has a structural lower bound
\begin{equation}
\dlts \approx  T^{3/2}_P \left( r\over c\right)^{1/3} =
{\Lambda^{2/3} r^{1/3} \over c}.
\label{eq:7}
\end{equation}
This lower bound is by many orders of magnitude smaller than the
\uc in the synchronization carried out by realistic quantum
clocks. But again, we shall rely only on the existence of
relation (\ref{eq:7}), the impossibility of actually reaching
the lower bound \dlts is not important.

\section{The Expression of the Uncertainties \dlt\ and \dlts\ in
Terms of Entities of the Future Unified Theory of General
Relativity and Quantum Theory}
\label{s:4}

Since in relations (\ref{eq:1}) and (\ref{eq:7}) \dlt and \dlts
are \ucs of the empty space-time, in the future unified theory
they will presumably take the form of vacuum spreads of
appropriate operators. The vacuum state $|V\rangle$ should
``know'' about general relativity, i.e.\ about gravitons, and if
the K model is sound, then in the non-relativistic approximation
\vr\  should represent the empty \st of \kar,  instead of the empty
Minkowskian space-time. As far as the appropriate operators are
concerned, they should refer to time because relations
(\ref{eq:1}) and (\ref{eq:7}) are time \uc relations, and since
these relations do not contain quantities describing particular
objects, the operators should not refer to particular objects
either. An effective local time operator \txt\  meets the above
requirements. We call this operator effective, because it may
well be that similarly to non-relativistic quantum mechanics,
there will be no time operator at the basic level of the unified
theory. 
The reason for considering the
operator \txt\  is that, as we shall see presently, the \ucs \dlt
and \dlts can be expressed in a simple way with the help of \txt\ 
and $|V\rangle$. Also, \txt\  exhibits two important features of the K
model. It states that time is not a global, but a local
quantity, in agreement with the involvement of general
relativity in the K model, and says that the values of the
moments of time have an uncertainty, corresponding to the \uc of the
\st structure.

We assume that the vacuum expectation value (the ``vev'') of
\txt\  is independent of $\mathbf x$ and equals $t$,
\begin{equation}
\left< \widehat t(\mathbf x, t) \right>_V = t,
\label{eq:8}
\end{equation} 
and we write $\widehat t$ in the convenient form
\begin{equation}
\txt = t + \taxt,
\label{eq:9}
\end{equation}
where, due to (\ref{eq:8}), 
\begin{equation}
\left< \taxt \right>_V = 0. 
\label{eq:10}
\end{equation}

Let us now look at relation (\ref{eq:1}). In this relation $T$
stands for the length of a time interval belonging to a segment
$\left[(\mathbf x, t), (\mathbf x, t')\right]$ of a $|\mathbf v
| = 0$ worldline in  the empty space-time  of \kar.  
It is therefore reasonable to identify $T$ with the vev of the
operator difference
\bee
\txtv - \txt.
\label{eq:11}
\eeq
From (\ref{eq:8}) and (\ref{eq:9}) we see that $T$ is equal to
the Minkowskian length $t' - t$ of our time interval (we take
$t' > t$):
\bee
T := \left< \txtv - \txt \right>_V = t' - t.
\label{eq:12}
\eeq

Since \dlt is the \uc of $T$ in the empty 
\kar\ space-time, it should be given by the
vacuum spread of the
operator difference in (\ref{eq:11}):
\bee
(\dlt)^2 := \left< (\txtv - \txt - T)^2 \right>_V . 
\label{eq:13}
\eeq
With equations (\ref{eq:9}) and (\ref{eq:12}) one finds that
\bee
(\dlt)^2 = \left< (\taxtv - \taxt )^2 \right>_V .
\label{eq:14}
\eeq
Relation (\ref{eq:1}) can be written now in the desired form
\bee
\left< (\taxtv - \taxt)^2 \right>_V \approx  T^{4/3}_P T^{2/3}.
\label{eq:15}
\eeq

We turn now to relation (\ref{eq:7}). There \dlts\ refers to the
relative time \uc at two world points on a $t =$ constant
hyperplane. \dlts\ should therefore be the vacuum spread of the
operator difference
\bee
\txvt - \txt = \taxvt - \taxt.
\label{eq:16}
\eeq 
With Equation (\ref{eq:8}) one finds that
\bee
\left< \txvt - \txt \right>_V = 0, 
\label{eq:17}
\eeq
therefore
\bee
(\Delta_L T)^2_{\rm syn} = \left< ( \taxvt - \taxt)^2 \right>_V,
\label{eq:18}
\eeq 
and relation (\ref{eq:7}) takes the form
\bee
\left< (\taxvt - \taxt)^2 \right>_V \approx  {T^{4/3}_P r^{2/3}
\over c^2},
\label{eq:19}
\eeq
where
\bee
r = |\mathbf x' - \mathbf x|.
\label{eq:20}
\eeq

A mathematical remark should be made here. The left-hand side of
relation (\ref{eq:19}) can be written in the form
\bee
\left< \widehat \tau^2(\mathbf x', t) + \widehat \tau^2 (\mathbf
x, t) - \taxvt \taxt - \taxt \taxvt \right>_V,
\label{eq:21}
\eeq
involving vev's of bilinear products of $\widehat\tau$'s taken
at equal time. It is well known from quantum field theory that
such vev's of local field operators are, as a rule, divergent,
and become finite only after renormalization. In the absence of
a detailed theory, one cannot evaluate the individual vev's in
(\ref{eq:21}). However, if relation (\ref{eq:7}) of the K model,
leading to (\ref{eq:19}), is correct, then the divergences in
(\ref{eq:21}) should cancel and the finite part should give the
right-hand side of (\ref{eq:19}). A similar remark applies to
relation~(\ref{eq:14}). 

\section{Uncertainties in the Phases of the Quantum\\
 States and
the Spread of the Relative Phases}
\label{s:5}

In regular non-relativistic quantum mechanics, the pure state of
an isolated physical system, constituted by $N$ microparticles
with masses $M_1$, $M_2$, \dots, $M_N$, is usually represented by a
\schr\ wave function
\bee
\Psi(x,t) = \exp \left(-{i \over \hbar} \widehat Ht \right) \cdot \Psi(x,0),
\label{eq:22}
\eeq
where $\widehat H$ stands for the Hamiltonian of the system, and
the evolution of $\Psi (x,t)$ takes place on the Minkowskian
space-time. Since the \schr\ evolution is deterministic, $\Psi$,
and consequently also its relative phases between any pairs of
points 
\bee
x = (\mathbf x_1, \dots, \mathbf x_N)
\label{eq:23}
\eeq
and $x'$ of the configuration space\footnote{%
Spin variables are omitted, because they do not play a role in
the K model.}
are sharply determined. We shall call such a wave function
``perfectly coherent''.

In the K model, the quantum state has to propagate on the \kar\
\st having the discussed \uc in its structure. As we have seen,
this \uc can be taken into account by substituting the effective
time operator \txt\ for the global Minkowskian time $t$:
\bee
t \to \txt = t + \taxt.
\label{eq:24}
\eeq

At this point we have to recall that on both sides of
Equation~(\ref{eq:22}) the rest energy phase factor
\bee
\exp(i \Phi(t)) := \exp \left( -{i \over \hbar} \sum^N_{\ell =
1} M_\ell c^2 t\right)
\label{eq:25}
\eeq
is omitted, because being independent of $x$ and of $p =
-i\hbar(\nabla_1, \dots, \nabla_N)$, it drops out from all the
observables. However, under the substitution (\ref{eq:24}), one
has to put in $\Phi(t)$
\bee
M_\ell c^2 t \to M_\ell c^2 \widehat t(\mathbf x_\ell, t),
\label{eq:26}
\eeq
because the coordinate of the $\ell$-th particle is $\mathbf
x_\ell$. This implies that
\bee
\Phi(t) \to \Phi(t) + \widehat \Phi(x,t),
\label{eq:27}
\eeq
where
\bee
\widehat \Phi(x,t) = - {c^2 \over \hbar} \sum^N_{\ell = 1}
M_\ell \widehat\tau (\mathbf x_\ell, t).
\label{eq:28}
\eeq
The rest energy phase $\Phi(t)$ can be omitted again, but the
$x$-dependent rest energy phase operator $\widehat \Phi(x,t)$ should be kept.

The substitution $t \to \widehat t(\mathbf x, t)$ should have
been carried out in the \schr\ wave function (\ref{eq:22}), too.
However, the non-relativistic matrix elements of $\widehat H$
are much smaller than the rest energies of the particles.
For solid bodies their contribution has been estimated in
\cite{2} in the framework of the \gmb\ model. In the present
paper we shall neglect it.

Keeping only the contribution of the rest energy phase, one
realizes that in the K model the \schr\ wave function acquires
an operator phase factor $\exp(i \widehat\Phi(x,t))$. In other
words, with any \schr\ wave function $\psi(x,t)$ one has to
associate a \kar\ state
\bee
\widehat\Psi_K (x,t) = \exp(i \widehat\Phi(x,t)) \cdot \Psi(x,t).
\label{eq:29}
\eeq
Unlike $\Psi(x,t)$, the K state $\widehat\Psi_K(x,t)$ is not
perfectly coherent. Through the operator phase factor it feels
the \uc of the space-time structure. Its departure from perfect
coherence between two points $x,x'$ of the configureation space
can be assessed by the amount of the \uc of its relative phase,
i.e.\ by the vacuum spread $\Delta_\Phi(x,x',t)$ of the relative
phase operator
\bee
\widehat\Phi_R(x,x',t) = \widehat \Phi(x',t) + \varphi(x',t) -
\widehat\Phi(x,t) - \varphi(x,t)
\label{eq:30}
\eeq
of the K state between those points. Here
\bee
\varphi(x,t) = \arg \Psi(x,t)
\label{eq:31}
\eeq
denotes the phase of $\Psi(x,t)$.

The $x,x'$ dependence of $\Delta_\Phi$ can be found  with the
help of relation (\ref{eq:19}). From Equations (\ref{eq:28}) and
(\ref{eq:10}) it follows that the vev of $\,\widehat\Phi_R$ is equal
to the \schr ian relative phase,
\bee
\left< \widehat\Phi_R(x,x',t) \right>_V = \varphi(x',t) - \varphi(x,t),
\label{eq:32}
\eeq
which drops then out from the vacuum spread of $\widehat\Phi_R$:
\begin{eqnarray}
\Delta^2_\Phi(x,x',t) & = & \left< (\widehat\Phi_R(x,x',t) -
\varphi(x', t) - \varphi(x,t) )^2 \right>_V \nonumber \\
& = & \left< (\widehat\Phi(x',t) - \widehat\Phi(x,t))^2 \right>_V .
\end{eqnarray} 
With Equation (\ref{eq:28}) one finds
\bee
\Delta^2_\Phi(x,x',t) = {c^4 \over\hbar^2} \sum^N_{i,\ell = 1} M_i
M_\ell \left< (\widehat\tau(\mathbf x'_i, t) - \widehat\tau(\mathbf x_i,
t)) (\widehat\tau(\mathbf x'_\ell, t) - \widehat \tau(x_\ell, t)
\right>_V .
\label{eq:34}
\eeq
The vev in the last expression is identically equal to
\begin{eqnarray}
& & {1\over 2} \left< \left( \widehat\tau(\mathbf x'_i, t) -
\widehat\tau(\mathbf x_\ell, t)\right)^2 + \left(\widehat\tau (\mathbf
x_i, t) - \widehat\tau(\mathbf x'_\ell, t) \right)^2 \right. \nonumber \\
& & \qquad  - \left. \left( \widehat\tau(\mathbf x'_i, t) - \widehat\tau(\mathbf
x'_\ell, t) \right)^2 - \left( \widehat \tau(\mathbf x_i, t) -
\widehat\tau(\mathbf x_\ell, t) \right)^2 \right>_V, 
\label{eq:35}
\end{eqnarray} 
and with relation (\ref{eq:19}) one obtains for the spread of
the relative phase the formula
\bee
\Delta^2_\Phi(x,x') \approx  \Lambda^{4/3} {c^2 \over \hbar^2}
\sum^N_{i,\ell = 1} M_i M_\ell\! \left(\! |\mathbf x'_i\! -\! \mathbf
x_\ell|^{2/3} - {1 \over 2} |\mathbf x_i \! - \! \mathbf x_\ell
|^{2/3} - {1 \over 2} |\mathbf x'_i\! -\! \mathbf x'_\ell|^{2/3}\!
\right)\!. 
\label{eq:36}
\eeq 
The time argument of $\Delta_\Phi$ has been omitted since
$\Delta_\Phi$ turned out to be time independent, although
$\widehat\Phi(x,t)$ may depend on~$t$.

According to Equation~(\ref{eq:36}), $\Delta_\Phi$ increases
with the masses and with the number of the microparticles
constituting the system, and for a given system it increases
with the distances $| \mathbf x'_i - \mathbf x_\ell|$, that is
with the separation between the points $x, x'$ in the
configuration space. These are encouraging features concerning
the expected loss of coherence between ``macroscopically
distinct'' components of the quantum state of a macroscopic body.

With the notations of the present paper, the formula for
$\Delta_\Phi$ derived in \cite{2} (see also \cite{5}) with the help of the
\gmb's reads 
\bee
\Delta^2_\Phi(x,x') \approx  \Lambda^{4/3} {c^2 \over\hbar^2} \int
{d^3 k \over k^{11/3}} |\mu_{\mathbf k}(x') - \mu_{\mathbf k}(x)
|^2, 
\label{eq:37}
\eeq
where
\bee
\mu_{\mathbf k}(x) = \sum_\ell M_\ell e^{i{\mathbf{kx}}_\ell}
\label{eq:38}
\eeq
is the Fourier transform of the mass distribution
\bee
\varrho(\mathbf x) = \sum^N_{\ell = 1} M_\ell \delta(\mathbf x -
\mathbf x_\ell)
\label{eq:39}
\eeq
of $N$ pointlike particles of masses $M_1, \dots, M_N$ in the
configuration $x = [\mathbf x_1, \dots, \mathbf x_N]$. From
Equation~(\ref{eq:37}) one sees that the \uc of the relative
phase increases with the difference (of the absolute values of
the Fourier transform) of the mass distribution of the $N$
particles in the configurations $x$ and $x'$.

The formula (\ref{eq:36}) for $\Delta_\Phi$ has been obtained
previously with the help of the $\{ t_\beta\}$ model by the
present author \cite{7}, who realized then that the Fourier integral
in the original formula (\ref{eq:37}) can be calculated
analytically and is equal to the sum in formula (\ref{eq:36}).
So, if
the expressions (\ref{eq:13}) and (\ref{eq:18}) for \dlt\ and
\dlts\ are reliable, then the influence of the single, quantum
mechanically uncertain \st on the relative phases has been
correctly mimicked by both classical \st models.

\section{Coherence Properties of the K states, Coherence Cells
and Cell Lengths}
\label{s:6}

\subsection{K States with Nearly Perfect and with Destroyed
Coherence. The Coherence Cell}
\label{ss:6.1}

One sees from formula (\ref{eq:37}) that for any pairs of points
$x' \neq x$, $\Delta_\Phi > 0$. This means that a K state $\widehat
\Psi_K$, normalized to $1$, is never perfectly coherent.

Concerning the norm of $\widehat\Psi_K$, from (\ref{eq:29}) one gets
\bee
\widehat\Psi^+_K(x,t) \widehat\Psi_K(x,t) = |\Psi(x,t)|^2,
\label{eq:40}
\eeq
because the unitary operator phase factor drops out from the
product. Therefore, the weight $w_\Omega$ of a K state in a
domain $\Omega$ of the configuration space is equal to the
weight of the \schr\ wave function associated with $\widehat\Psi_K$,
\bee
w_\Omega = \int_\Omega dx | \Psi(x,t)|^2,
\label{eq:41}
\eeq
and $\widehat\Psi_K$ is normalized to $1$ together with $\Psi$.
(Pedantically, in (\ref{eq:40}) one should consider the vev of
$\;\widehat\Psi^+_K(x,t) \widehat\Psi_K(x,t)$, but it is obviously equal
to the product itself.)

Let us now see how the coherence of the K states can be
characterized. 

\smallskip
(1) If $\widehat\Psi_K$ occupies\footnote{%
As a rule, $\Psi$ is different from zero almost everywhere.
Therefore, strictly speaking, $\Psi$, and then $\widehat\Psi_K$ too,
occupy the whole configuration space. In our terminology the
``domain occupied by $\Psi$'' is the smallest domain in which
the weight $w$ of $\Psi$ is close to the maximal weight $1$
(e.g.\ $w = 1 - 10^{-4}$). The expansion and the shrinking (or
the localization) of $\Psi$ means that this domain expands or
shrinks.} 
a domain $\Omega$ of the configuration space such that
\bee
\Delta_\Phi(x,x') \ll \pi \ \mbox{ for all } \ x,x' \in \Omega,
\label{eq:42}
\eeq
then in good approximation the \ucs of the relative phases of
$\widehat\Psi_K$ can be neglected, and the relative phases of
$\widehat\Psi_K$ are practically equal to those of $\Psi$. In such a
case we shall say that the coherence of $\widehat\Psi_K$ is nearly
perfect. 

\smallskip
(2) If $\widehat \Psi_K$ occupies a domain containing
non-overlapping subdomains $\Omega$, $\Omega'$ such that
\bee
\Delta_\Phi(x,x') \geqslant \pi \ \mbox{ for all } \ x\in \Omega 
\ \mbox{ and all } \ x' \in \Omega',
\label{eq:43}
\eeq
then the relative phases of $\widehat\Psi_K$ between these
subdomains are completely uncertain, the coherence between the
components of $\widehat\Psi_K$ belonging to these subdomains is
destroyed. Notice, however, that within smaller subdomains
inside which $\Delta_\Phi < \pi$, a certain degree of coherence
persists, and inside sufficiently small subdomains even
$\Delta_\Phi \ll \pi$ holds, so that within such a small domain
$\widehat \Psi_K$ is near to perfect coherence. 

The maximal domains $\Omega_c$ of the configuration space such that
\bee
\Delta_\Phi(x,x') \leqslant \pi \ \mbox{ for all } x,x' \in \Omega_c,
\label{eq:44}
\eeq
have been called ``coherence cells'' in \cite{2}. A K state occupying
a single cell is still not incoherent, but if it occupies
non-overlapping cells, then it is incoherent (it has incoherent
components of non-negligible weights).

The size and the shape of the coherence cell depends strongly on
the composition of the physical system considered. Below we
shall look at the cells of microparticles and of a class of
solid objects. As we shall see, the cells of these systems are
spherical, and can therefore be characterized by a single
parameter, the diameter of the cell.

\subsection{The Coherence Cell and the Cell Length of a Single
Microparticle} 
\label{ss:6.2}

For a single microparticle of mass $M = M_1$, one easily finds
from formula (\ref{eq:36}) that
\bee
\Delta_\Phi(a) \approx  {\Lambda^{2/3} \over L} a^{1/3}, 
\label{eq:45}
\eeq
where (with $\mathbf x = \mathbf x_1$)
\bee
a = |\mathbf x' - \mathbf x|,
\label{eq:46}
\eeq
and
\bee
L = {\hbar \over Mc}
\label{eq:47}
\eeq
is the Compton wavelength of the particle. Notice that
$\Delta_\Phi$ increases monotonically with the distance $a$. The
coherence cell is therefore a sphere of diameter $a_c$ in the
configuration space $x = (\mathbf x)$, where $a_c$ is equal to
the value of $a$ at which $\Delta_\Phi$ reaches the value $\pi$:
\bee
\Delta_\Phi(a_c) = \pi \approx  1.
\label{eq:48}
\eeq
From (\ref{eq:45}) one finds that
\bee
a_c \approx  \left( L \over \Lambda \right)^2 L.
\label{eq:49}
\eeq
So, for a microparticle the coherence cell is characterized by a
single parameter, the cell length $a_c $. In \cite{2} the term ``cell
diameter'' has been used. However, there are physical systems
the coherence cell of which is not spherical, and then $a_c$ is
not a diameter.

For the electron $L \approx  10^{-11}$~cm, and
\bee
a_c \approx  10^{33}~{\rm cm}.
\label{eq:50}
\eeq
The cell length of the electron, and also of the other
microparticles, has a supraastronomical value. Therefore, any
realistic \schr\ wave function $\Psi$ of an isolated
microparticle occupies only a tiny part of a coherence cell, a
part inside which $\Delta_\Phi \lll \pi$, and the K state
associated with $\Psi$ is always practically perfectly coherent.
Due to the very small masses of the microparticles, the \uc of
the \st structure has no appreciable effect on a single
microparticle. For the same reason this is also true for any
isolated microsystem (for a system consisting of a few
microparticles, free or bound). 

\subsection{The Coherence Cell and the Cell Length of Spherical,
Homogeneous Solid Objects}
\label{ss:6.3}

In a homogeneous object there are $N \gg 1$ identical
microscopic constituents (e.g.\ molecules). Formula
(\ref{eq:36}) for $\Delta_\Phi$ then becomes
\bee
\Delta^2_\Phi(x,x') \approx  {\Lambda^{4/3} \over L^2_{\rm
micro}} \sum^N_{i,\ell = 1} \left( |\mathbf x'_i - \mathbf
x_\ell |^{2/3} - {1 \over 2} |\mathbf x_i - \mathbf
x_\ell|^{2/3} - {1\over 2} |\mathbf x'_i - \mathbf
x'_\ell|^{2/3} \right),
\label{eq:51}
\eeq
where
\bee
L_{\rm micro} = {\hbar \over  M_{\rm micro}\, c}
\label{eq:52}
\eeq
is the Compton wavelength of a constituent of mass $M_{\rm micro}$.

In a solid object the constituents are at, or very close to,
their equilibrium positions. Consequently, the  \schr\ wave
function of an isolated solid object is practically zero
everywhere, except in such points $x = (\mathbf x_1, \dots,
\mathbf x_N)$ of the configuration space, in which the
constituents are at their equilibrium positions. Therefore, in
the case of a homogeneous, spherical solid object of radius $R$,
in $\Delta_\Phi(x,x')$ the $\mathbf x_i$ coordinates belonging
to $x$ are distributed uniformly in the volume of a sphere of
radius $R$, and the $\mathbf x'$ coordinates belonging to $x'$
fill uniformly an other such sphere in the three dimensional
$XYZ$ space. 

For a solid object of arbitrary shape, two equilibrium
configurations $x, x'$ differ from each other by a translation
of their center of mass (c.m.) and by a rotation leaving the
c.m.\ fixed. For a spherical homogeneous object,
$\Delta_\Phi(x,x')$ does not change appreciably under a
rotation, because $\Delta_\Phi$ is invariant under any
permutation of the $\mathbf x$ coordinates among themselves and
of the $\mathbf x'$ coordinates among themselves. Therefore, we
have to consider only such configurations which differ from each
other by a translation
\bee
\mathbf x'_i = \mathbf x_i + \hbox{\boldmath$a$\unboldmath},
\label{eq:53}
\eeq
where
\bee
\hbox{\boldmath$a$\unboldmath} = \mathbf x'_{\rm c.m.} - \mathbf x_{\rm c.m.}
\label{eq:54}
\eeq
is the vector joining the centers of the spheres, which are also
the centers of mass of the objects in the two configurations.

With (\ref{eq:53}) the sum $\sum$ in (\ref{eq:51}) becomes
\bee
\sum = \sum^N_{i,\ell = 1} \left( |\mathbf x_i - \mathbf x_\ell
+ \hbox{\boldmath$a$\unboldmath}|^{2/3} - | \mathbf x_i - \mathbf x_\ell|^{2/3} \right).
\label{eq:55}
\eeq
The $\mathbf x_i$ and the $\mathbf x_\ell$ coordinates fill a
sphere uniformly, $N \gg 1$, and the expression under the sum is
a continuous, slowly varying function of the coordinates.
Therefore, the double sum in (\ref{eq:55}) can be replaced, in a
good approximation, by a double integral. With $\mathbf x_i \to
\mathbf r$, $\mathbf x_\ell \to \mathbf r'$, one gets
\bee
\sum = {N^2 \over V^2} \int_V d^3 r \int_V d^3 r'\left(| \mathbf r -
\mathbf r' + \hbox{\boldmath$a$\unboldmath} |^{2/3} - | \mathbf r - \mathbf r'|^{2/3}
\right),
\label{eq:56}
\eeq
where the integrals have to be taken over the volume of the same
sphere of radius $R$. $\sum$ can be calculated for any value of
{\boldmath$a$\unboldmath}, but it is more enlightening to evaluate it in two
extreme situations, namely when $|\hbox{\boldmath$a$\unboldmath}| \equiv a \gg R$ and
when $a \ll R$.

a) The case $a \gg R$

In this case one has also $a \gg |\mathbf r - \mathbf r'|$,
because in (\ref{eq:56}) $|\mathbf r - \mathbf r'| \leqslant 2R$.
Therefore, in good approximation,
\bee
|\mathbf r - \mathbf r' + \hbox{\boldmath$a$\unboldmath}|^{2/3} - |\mathbf r -
\mathbf r'|^{2/3} = a^{2/3},
\label{eq:57}
\eeq
and
\bee
\sum = N^2 a^{2/3}. 
\label{eq:58}
\eeq
Noticing that
\bee
{N \over L_{\rm micro}} = {1 \over L},
\label{eq:59}
\eeq
where $L$ is the Compton wavelength correspondig to the mass $M
= NM_{\rm micro}$ of the object considered,
Equation~(\ref{eq:51}) becomes
\bee
\Delta_\Phi(x,x') \approx  \Lambda^{2/3} {a^{1/3} \over L}, \qquad
a \gg R.
\label{eq:60}
\eeq
This formula is formally identical with the one obtained for a
single microparticle, but $a$ stands now for the distance
between the centers of mass of the object in the two
configurations $x,x'$. The coherence cell is again a sphere of diameter
\bee
a_c \approx  \left( L \over \Lambda \right)^2 L, \qquad a_c \gg R,
\label{eq:61}
\eeq
but now in the center of mass coordinate subspace of the
configuration space.

\smallskip
b) The case $a \ll R$

In this case in (\ref{eq:56}) $a \ll |\mathbf r - \mathbf r'|$,
except for a small subdomain of the integration domain. The
detailed calculation shows that one can forget about the
violation of the condition $a \ll |\mathbf r - \mathbf r'|$.
Expanding $| \mathbf r - \mathbf r' + \hbox{\boldmath$a$\unboldmath}|^{2/3}$ in
powers of $a$, one finds that the 
leading contribution to $\sum$ is of the order $a^2$. Since the
dimension of $\sum$ is $cm^{3/2}$, and apart from $a$ the only
length parameter in (\ref{eq:56}) is $R$, one finds that
\bee
\sum \approx  N^2 {a^2 \over R^{4/3}}, \qquad a \ll R.
\label{eq:62}
\eeq
Equation (\ref{eq:51}) takes now the form
\bee
\Delta_\Phi(x,x') \approx  \left( \Lambda \over R \right)^{2/3} {a
\over L}, \qquad a \ll R.
\label{eq:63}
\eeq
$\Delta_\Phi$ increases again monotonically with $a$, and
reaches the value $\pi \approx  1$ when $a$ is equal to 
\bee
a_c \approx  \left( R \over \Lambda \right)^{2/3} L,
\qquad a_c \ll R.
\label{eq:64}
\eeq

Formulas (\ref{eq:61}) and (\ref{eq:64})  for the cell length of
spherical, homogeneous objects have been presented in \cite{2} as
``the most important results'' of the model. Their physical
meaning has been discussed in due detail in \cite{2, 4, 3}. Here we
recall only that $a_c \gg R$ is the region where quantum
behavior,
\hbox{$a_c \ll R$} --- the region where classical behavior
dominates. Indeed, if $a_c \gg R$, then the \schr\ wave function
$\Psi (x,t)$ may have an \uc of the order of $a_c$ in the
position of the c.m., much larger than the geometrical size $R$
of the object, without making the K state associated with $\Psi$
incoherent. Large coherent \ucs correspond to quantum
behavior. On the contrary, when $a_c \ll R$, the $K$ state
becomes incoherent when the \uc of the position of the c.m.\ is
still much smaller than the geometrical size $R$ of the object.
In other words, when $a_c \ll R$, the coherent, quantum
mechanical \uc of the c.m.\ is much smaller than $R$. Small
quantum mechanical positional \uc is a characteristic of
classical behavior. 

\subsection{Tiny Grains, Macroscopic Bodies and the Transition
Region between Quantum and Classical Behavior}
\label{ss:6.4}

It can be shown that formulas (\ref{eq:61}) and (\ref{eq:64})
hold, with small corrections absorbable in the symbol $\approx $,
in the whole region $a_c \geqslant R$ and $a_c \leqslant R$,
respectively. In particular, they hold also in the region $a_c
\approx  R$. 
It follows from the discussion in the preceding subsection that
the latter region is, for spherical homogeneous objects, the
transition region between quantum and classical behavior. In
this region the maximal quantum mechanical \uc of the position
of the c.m.\ is of the same order of magnitude as the
geometrical size of the object. For usual terrestrial densities
$\varrho \approx  1\gcm$, one easily finds from Equation
(\ref{eq:61}) (as well as from (\ref{eq:64}), of course) that
\begin{eqnarray}
a^{\rm tr}_c & \approx  & R^{\rm tr} \approx 
10^{-5}~{\rm cm}, \label{eq:65} \\
M^{\rm tr} & = & {4\pi \over 3} \varrho(R^{\rm
tr})^3 \approx  10^{-14}~{\rm g}. \label{eq:66}
\end{eqnarray}
Thus, the transition mass region is the region of the colloidal
grains and dust particles. The region $a_c \gg R$ of quantum
behavior corresponds to tiny grains with masses much smaller
than $M^{\rm tr}$. To give an example, for a grain of $R
\approx  10^{-6}$~cm and of mass $M \approx  \varrho R^3 \approx 
10^{-18}$~g, one finds\footnote{%
The estimate $a_c \approx  10$~km, quoted in \cite{3}, is erroneous.}
from (\ref{eq:61}) that $a_c \approx  10^5$~km, indeed much larger
than $R$. If the wave function of such an isolated grain
expanded, say, over $10^3$~km in the c.m.\ coordinate subspace,
its associated K state would remain still very coherent. On the
contrary, for a ball of $M \approx  1$~g and radius $R \approx 
1$~cm, one finds from (\ref{eq:64}) that
\bee
a_c \approx  10^{-16}~{\rm cm},
\label{eq:67}
\eeq 
a value much smaller indeed than $R$. The K model states that
two positions of this ball, with separation $2 a_c \approx  2
\cdot 10^{-16}$~cm between the centers of mass, correspond
already to an incoherent K state, these positions are already
``macroscopically distinct''.
We are in the region of predominantly classical behavior, with
$a_c \ll R$, $M \gg M^{\rm tr}$. It should be noted that formula
(\ref{eq:64}) for $a_c$ is applicable only to bodies of moderate
size. Above $R = 1$~m the elastic vibrations of the body are not
negligible~\cite{2}.

\section{The Stochastic Modification of the \schr\ Evolution of
Isolated Systems}
\label{s:7}

The mode of the combination of a stochastic process with the
\schr\ evolution is suggested by the behavior of the K state
\bee
\widehat\Psi_K(x,t) = \exp(i\widehat\Phi(x,t)) \cdot \Psi(x,t)
\label{eq:68}
\eeq
during the \schr\ evolution of its associated $\Psi$ function.
As it is well known, during that evolution $\Psi$ expands\footnote{%
For appropriately awkward initial wave functions, a transient
shrinking precedes the steady expansion.}, 
at least in the c.m.\ subspace of the configuration space.
According to Equations (\ref{eq:68}) and (\ref{eq:40}),
$\widehat\Psi_K$ expands (or shrinks) exactly in the same way as
$\Psi$. However, while $\Psi$ remains perfectly coherent, the
coherence of $\widehat\Psi_K$ deteriorates during the expansion, and
when it occupies a domain larger than a coherence cell, it has
already incoherent components.

The basic idea \cite{1, 2} for the introduction of the stochastic
element is that the K state $\widehat\Psi_K$ counterbalances its
loss of coherence by stochastically and instantaneously
localizing itself, of course together with $\Psi$, to one of the
coherence cells lying in the domain to which the state expanded.
After a localization, $\Psi$ expands again under the \schr\
equation, until the situation gets ripe for a new localization,
and so on.

There are many, essentially equivalent ways to specify these
expansion--localization cycles. In \cite{2}, for an isolated  system
the coherence cell of which is characterized by a single cell
diameter $a_c$, an instantaneous random localization occurs
whenever the quantum state occupies a domain of diameter $2a_c$.
The localizations occur then at discrete moments of time, at
intervals 
\bee
\tau_c \approx  {M a^2_c \over \hbar}, 
\label{eq:69}
\eeq
the time needed for the \schr\ wave function of an isolated
system to expand from size $a_c$ to $2a_c$.

It has been shown, among others by Ghirardi, Rimini and Weber
\cite{10} and by Caves and Milburn \cite{11}, that the stochastic
localizations from size $2a_c$ to $a_c$ occurring at equally
spaced discrete moments of time, can be replaced by
infinitesimal stochastic localizations from size $a_c + da_c$ to
$a_c$ occurring continuously in time, blended with the
continuous \schr\ evolution. The application of the
infinitesimal GRW localizations to the K model,
in the case of a single cell length $a_c$,
 has been carried
out in \cite{9}.

Let us now look at the expansion--localization cycles of various
physical systems. Since for the electron $a_c \approx 
10^{33}$~cm, a realistic wave function of an electron occupies
only a tiny part of a coherence cell. Therefore, the \schr\
evolution is not interrupted by localizations. However, if a
ball of 1~g could be isolated, then as soon as its $\Psi$ function
would expand to a domain of diameter $2a_c \approx  2 \cdot
10^{-16}$~cm (in the c.m.\ subspace), a stochastic localization 
onto a single cell of diameter $a_c$ would take place. So,
according to the K model, the localization of the c.m.\ of the
ball would remain practically pointlike even if the ball were
isolated. Notice that from $\Delta x \cdot \Delta v \approx  \hbar
/ M$ one finds $\Delta v \approx  10^{-11}$~cm/sec. The \uc of the
velocity would be very small, too.

Of course, a macroscopic ball cannot be perfectly isolated. The
interplay of the \kar\ law of time evolution with the effects
caused by the surroundings has been discussed in \cite{2, 4, 9}.

\kar\ suceeded in applying the basic idea of his time evolution
law to systems of many 
quasi independent
degrees of freedom. In particular, he
worked out the process of the decay of the superposition of
tracks in a cloud chamber. The detailed discussion is  given
only in \cite{2}. For a short outline, see \cite{5, 6, 3}.

\section{Concluding Remarks}
\label{s:8}

The expression of the structural \ucs \dlt\ and \dlts\ through
$\widehat\tau$ and \vr\ and the derivation of the formula for the
spread $\Delta_\Phi$ of the relative phases without the use of a
\st model, is a tentative step towards turning the K model into a theory.
A further step could concern the law of time evolution described
in the preceding section. There, the phase operator
$\widehat\Phi$ producing the spread of 
the relative phases acts as  a ``hop-master''. It tells when
$\Psi$, and with it $\widehat\Psi_K$, should jump in order to
prevent the loss of coherence of $\widehat\Psi_K$. It would be much
better if one could set up a general stochastic differential
equation for $\widehat\Psi_K$, presumably of It\^o type, and let the
equation work. However, this does not seem to be possible without
knowing more about the phase operator $\widehat\Phi$, that is about
the local time operator \taxt\ entering into $\widehat\Phi$. The
basic relations (\ref{eq:1}) and (\ref{eq:7}) made possible only
to evaluate the vacuum spreads (\ref{eq:15}) and (\ref{eq:19}),
involving $\widehat\tau$'s. This was sufficient to derive the
formulas for $\Delta_\Phi$ and then for $a_c$, and to formulate
with their help the hop-master's rule. Of course, it is
possible, after having calculated the cell length $a_c$ (or the
cell lengths $a^{(1)}_c, a^{(2)}_c, \dots)$ of a given system,
to set up an It\^o equation containing these $a_c$'s, which
would smoothly reproduce the results of the bumpy hop-master's
rule. What we have in mind is a general equation for $\widehat
\Psi_K$, not containing the parameters $a^{(i)}_c$ of the
particular system to which the equation is applied. It is
probable that such an equation cannot be found without knowing
more about the unified theory of general relativity and quantum
theory. 

Another open probelm is the relativistic generalization of the K
model. This is a common open problem of the existing models with
stochastic modification of the \schr\ evolution. In the case of
the K model, a specific task arises. One should find, first of
all, a covariant description for the structural \uc of the \kar\
space-time. Again, there is little hope for progress without the
unified theory.

A direction into which progress can be made is the derivation of
the basic relation (\ref{eq:1}) between $T$ and \dlt. This
relation has been deduced by \kar\ in three different ways \cite{1,
2}; \cite{2, 5}; \cite{4}. On the one hand, it is reassuring that various
approaches lead to the same result. On the other hand, each of
these approaches, taken separately, has loose ends. For
instance, in \cite{2} and \cite{5} relation (\ref{eq:1}) has been
obtained from the study of the quantum behavior of the hand of a
clock, and one can legitimately ask what would happen if the
dial also entered the game. The answer \cite{12} is that the dial 
would not upset relation (\ref{eq:1}). This will be shown in a
forthcoming paper, where  the derivation(s) of relation
(\ref{eq:1}) will be scrutinized and a comparison of \kar's
clock with a Wigner--Salecker clock \cite{13} will also be made.

Relations (\ref{eq:1}) and (\ref{eq:7}) have been recently
rediscovered by Ng and Van Dam \cite{14}. These authors obtain also
the formula for the spread of the relative phases for a single
particle (our formula (\ref{eq:45}), but they do not consider
composite systems. Having only $\Lambda$, $a$ and $L$ at their
disposal, they had to choose $(\Delta_\Phi \approx  1$, $a
\lesssim L)$ as the condition for classical behavior, which
leads then to the condition $m \gtrsim m_P \approx  10^{-5}$~g,
$m_P$ being the Planck mass. The authors remark that this
condition is only a sufficient condition for classical behavior.
Indeed, objects with masses much smaller than $10^{-5}$~g are
known to behave classically. As we have seen, in the K model the
transition mass depends not  only on $m_P$ (i.e.\ on $\Lambda$),
and for homogeneous spherical solid objects the sufficient and
necessary condition for classical behavior is 
$a_c \ll R$.
 It should be noted that in \cite{14}
there are many interesting relations, not overlapping with 
the K model.

The author is indebted to F. \kar\ for countless discussions
scattered over the last three decades.
This work was partially
supported by the Hungarian Research Fund, under grants OTKA
T016246 and OTKA T030374.

\end{document}